\documentclass[]{interact}

\usepackage{epstopdf}
\usepackage{subfigure}
\usepackage{multirow}
\usepackage{float}
\usepackage[numbers,sort&compress]{natbib}
\bibpunct[, ]{[}{]}{,}{n}{,}{,}
\newcommand{\f}{\operatorname}

\theoremstyle{plain}

\theoremstyle{definition}

\theoremstyle{remark}

\begin{document}

\articletype{}

\title{Some Computational Aspects to Find Accurate Estimates  for the Parameters of the Generalized Gamma distribution }

\author{ Jorge Alberto Achcar$^{\rm a}$, Pedro Luiz Ramos$^{\rm b}$$^{\ast}$\thanks{$^\ast$Corresponding author. Email: pedrolramos@usp.br
\vspace{6pt}}, and Edson Zangiacomi Martinez$^{\rm a}$ \\\vspace{6pt} 
$^{a}${Ribeir\~ao Preto School of Medicine, University of S\~ao Paulo, Ribeir\~ao Preto, Brazil} \\
 $^{b}${Institute of Mathematical Science and Computing, University of S\~ao Paulo, S\~ao Carlos, Brazil} }

\maketitle

\begin{abstract}
In this paper, we discuss computational aspects to obtain accurate inferences for the parameters of the generalized gamma (GG) distribution. Usually, the solution of the maximum likelihood estimators (MLE) for the GG distribution have no stable behavior depending on large sample sizes and good initial values to be used in the iterative numerical algorithms. From a Bayesian approach, this problem remains, but now related to the choice of prior distributions for the parameters of this model. We presented some exploratory techniques to obtain good initial values to be used in the iterative procedures and also to elicited appropriate informative priors. Finally, our proposed methodology is also considered for data sets in the presence of censorship.
\end{abstract}

\begin{keywords}
Bayesian Inference; Classical inference; Generalized gamma distribution; Random censoring.
\end{keywords}

\section{Introduction}\label{sec:1}

The Generalized gamma (GG) distribution is very flexible to be fitted by reliability data due to its different forms for the hazard function. This distribution was introduced by Stacy \cite{stacy1962} and has probability density function (p.d.f.) given by
\begin{equation}\label{denspgg}
f(t|\boldsymbol{\theta})= \frac{\alpha}{\Gamma(\phi)}\mu^{\alpha\phi}t^{\alpha\phi-1}\exp\left(-(\mu t)^{\alpha}\right) ,
\end{equation}
where $t>0$ and $\boldsymbol{\theta}=(\phi,\mu,\alpha)$ is the vector of unknown parameters. The parameters $\alpha>0$ and $\phi >0$ are shape parameters and $\mu >0$ is a scale parameter. This parametrization is a standard form of the GG distribution originated from a general family of survival functions with a location and scale changes of the form, $Y=\log(T)= \mu +\sigma W$ where $W$ has a generalized extreme value distribution with parameter $\phi$ and  $\alpha=\frac{1}{\sigma}$. The survival function is given by
\begin{equation*}
S(t|\boldsymbol{\theta})= \int_{(\mu t)^{\alpha}}^{\infty}{\frac{1}{\Gamma(\phi)}w^{\phi -1}e^{-w}}dw =\frac{\gamma\left[\phi,(\mu t)^{\alpha}\right]}{\Gamma(\phi)}\ ,
\end{equation*}
where $\Gamma[y,x]=\int_{x}^{\infty}{w^{y-1}e^{-w}dw}$ is the upper incomplete gamma function. Many lifetime distributions are special cases of the GG distribution such as the Weibull distribution (when $\phi=1$), the gamma distribution ($\alpha=1$), log-Normal (limit case when $\phi \rightarrow \infty$) and the generalized normal distribution ($\alpha=2$). The generalized normal distribution also is a distribution that includes various known distributions such as half-normal ($\phi=1/2,\mu=1/\sqrt{2}\sigma$), Rayleigh ($\phi=1,\mu=1/\sqrt{2}\sigma$), Maxwell-Boltzmann ($\phi=3/2$) and chi ($\phi=k/2 , k=1,2,\ldots$).

Although the inferential procedures for the gamma distribution can be easily be obtained by classical and Bayesian approaches, especially using the computational advances of last years in terms of hardware and software \cite{antonio2013bayesian}, the inferential procedures for the generalized gamma distribution under the maximum likelihood estimates (MLEs) can be unstable (e.g., \cite{stacy1965parameter, parr1965method, hager1970inferential, wingo1987computing}) and its results may depend on the initial values of the parameters chosen in the iterative methods. Ramos et al. \cite{ramos2014metodo} simplified the MLEs under complete samples and used the obtained results as initial values for censored data. However, the obtained MLEs may not be unique returning more than one root. Additionally, the obtained results lies under asymptotic properties in which are not achieved even for large samples ($n>400$) as was discussed by Prentice \cite{prentice1974log}. A similar problem appears under the Bayesian approach, where different prior distributions can have a great effect on the subsequent estimates. It is important to point out that, the literature on  GG distribution is very extensive, with many studies  considering statistical inference as well as others more concentrated in applications \cite{cox2007parametric,  shin2005statistical, diciccio1987approximate, pham1995generalized, de2011kumaraswamy, ashkar1988generalized, ramoslouzada2016, ramos2016efficient}.
	
In this study, methods exploring modifications of Jeffreys' priors to obtain initial values for the parameter estimation are proposed. The obtained equations under complete and censored data were used as initial values to start the iterative numerical procedures as well as to elicited informative prior under the Bayesian approach. An intensive simulation study is presented in order to verify our proposed methodology. These results are of great practical interest since it enable us for the use of the GG distribution in many application areas. The proposed methodology is illustrated in two data sets. 

This paper is organized as follows: Section 2 reviews the MLE for the GG distribution. Section 3 presents the Bayesian analysis. Section 4 introduces the methods to obtain good initial values under complete and censored data. Section 5 describes a simulation study to compare both classical and Bayesian approaches. Section 6 presents an analysis of the data set. Some final comments are made in Section 7.

\section{Classical inference}\label{sec:2}

Maximum likelihood estimates are obtained from from the maximization of the likelihood function . Let $T_1,\ldots,T_n$ be a random sample of a GG distribution, the likelihood function is given by
\begin{equation}\label{verogg1} 
L(\boldsymbol{\theta};\boldsymbol{t})=\frac{\alpha^n}{\f \Gamma(\phi)^n}\mu^{n\alpha\phi}\left\{\prod_{i=1}^n{t_i^{\alpha\phi-1}}\right\}\exp\left\{-\mu^{\alpha}\sum_{i=1}^n t_i^\alpha\right\}. \end{equation}

From $\frac{\partial}{\partial \alpha}\log(L(\boldsymbol{\theta};\boldsymbol{t}))$, $\frac{\partial}{\partial \mu}\log(L(\boldsymbol{\theta};\boldsymbol{t}))$ and $\frac{\partial}{\partial \phi}\log(L(\boldsymbol{\theta};\boldsymbol{t}))$ equal to zero, the obtained likelihood equations are
\begin{equation*} n\f \psi(\hat\phi)=n\hat\alpha\log(\hat\mu)+\hat{\alpha}\sum_{i=1}^n\log(t_i) \end{equation*}
\begin{equation}\label{verogg22}  n\hat\phi={\hat\mu}^{\hat\alpha}\sum_{i=1}^n{t_i}^{\hat\alpha} \end{equation}
\begin{equation*} \dfrac{n}{\hat{\alpha}}+n\hat\phi\log(\mu)+\phi\sum_{i=1}^n\log(t_i)=\hat\mu^{\hat\alpha}\sum_{i=1}^n{t_i}^{\hat\alpha}\log(\hat\mu t_i), \end{equation*}
where $\psi(k)=\frac{\partial}{\partial k}\log\Gamma(k)=\frac{\Gamma'(k)}{\Gamma(k)}$. The solutions of (\ref{verogg22}) provide the maximum likelihood estimates \cite{stacy1965parameter, hager1970inferential}. Numerical methods such as Newton-Rapshon are required to find the solution of the nonlinear system.

Under mild conditions the maximum likelihood estimators of $\boldsymbol{\theta}$ are not biased and asymptotically efficient. These estimators have an asymptotically normal joint distribution given by
\begin{equation*} 
(\boldsymbol{\hat{\theta}}) \sim N_k[(\boldsymbol{\theta}),I^{-1}(\boldsymbol{\theta})] \mbox{ for } n \to \infty , 
\end{equation*}
where $I(\boldsymbol{\theta})$ is the Fisher information matrix
\begin{equation*}
I(\boldsymbol{\theta})=
\begin{bmatrix}
 \dfrac{1+2\psi(\phi)+\phi\psi ' (\phi)+\phi\psi(\phi)^2}{\alpha^2} & -\dfrac{1+\phi\psi(\phi)}{\mu} & -\dfrac{\psi(\phi)}{\alpha} \\
 -\dfrac{1+\phi\psi(\phi)}{\mu} & \dfrac{\phi\alpha^2}{\mu^2}  & \dfrac{\alpha}{\mu} \\
 -\dfrac{\psi(\phi)}{\alpha} & \dfrac{\alpha}{\mu} & \psi ' (\phi)
\end{bmatrix} .
\end{equation*}

In the presence of censored observations, the likelihood function is
\begin{equation}\label{verossccens} L(\boldsymbol{\theta};\boldsymbol{t},\boldsymbol{\delta})=\frac{\alpha^d\mu^{d\alpha\phi}}{\f \Gamma(\phi)^n}\left\{\prod_{i=1}^nt_i^{\delta_i\alpha\phi-1}\right\}\exp\left\{-\mu^\alpha\sum_{i=1}^n\delta_i{t_i}^\alpha\right\}\prod_{i=1}^n(\f \Gamma[\phi,(\mu t_i)^\alpha])^{1-\delta_i}. \end{equation}

From $\frac{\partial}{\partial \alpha}\log(L(\boldsymbol{\theta};\boldsymbol{t},\boldsymbol{\delta}))$, $\frac{\partial}{\partial \mu}\log(L(\boldsymbol{\theta};\boldsymbol{t},\boldsymbol{\delta}))$ and $\frac{\partial}{\partial \phi}\log(L(\boldsymbol{\theta};\boldsymbol{t},\boldsymbol{\delta}))$ equal to zero, we get the likelihood equations,
\begin{equation*}
\begin{aligned}
\sum_{i=1}^{n}&\left\{(1-\delta_i)\left[\frac{(\mu t_i)^{\alpha\phi}e^{-(\mu t_i)^{\alpha}}\log(\mu t_i)}{\Gamma\left[\phi,\mu t_i^\alpha\right]}\right] \right\} = \frac{d}{\alpha}+d\phi\log(\mu)\\ & \ \ \ \ \ \ \ \ \ +\phi\sum_{i=1}^{n}\delta_i\log(t_i) -\mu^\alpha\sum_{i=1}^{n}\delta_i t_i^{\alpha}\log(\mu t_i)
\end{aligned}
\end{equation*}
\begin{equation}\label{eqvercen2}  
\frac{d\alpha\phi}{\mu}-\alpha\mu^{\alpha-1}\sum_{i=1}^{n}{\delta_i t_i^\alpha}=\sum_{i=1}^{n}\left\{(1-\delta_i)\left[\frac{\alpha t_i(\mu t_i)^{\alpha\phi-1}e^{-(\mu t_i)^{\alpha}}}{\Gamma\left[\phi,\mu t_i^\alpha\right]}\right] \right\}
\end{equation}
\begin{equation*} 
d\alpha\log(\mu)-n\psi(\phi)+\alpha\sum_{i=1}^{n}{\delta_i\log(t_i)}= -\sum_{i=1}^{n}\left\{(1-\delta_i)\left[\frac{\Psi\left[\phi,\mu t_i^\alpha\right]}{\Gamma\left[\phi,\mu t_i^\alpha\right]}\right] \right\}
\end{equation*}
where $\Psi(k,x)=\frac{\partial}{\partial k}\f \Gamma[k,x]=\int\limits\limits_x^{\infty}w^{k-1}\log(w)e^{-w}dw$. The solutions of the above non-linear equations provide the MLEs of $\phi$, $\mu$ and $\alpha$. Other methods using the classical approach have been proposed in the literature to obtain inferences on the parameters of the GG distribution \cite{huang2006new, diciccio1987approximate, ahsanullah2013kernel}.

\section{A Bayesian approach}\label{sec:3}

The GG distribution has nonnegative real parameters. A joint prior distribution for $\phi$, $\mu$ and $\alpha$ can be obtained from the product of gamma distributions. This prior distribution is given by
\begin{equation}\label{pgammagg}
\pi_G\left(\phi,\mu,\alpha\right)\propto \phi^{a_1-1}\mu^{a_2-1}\alpha^{a_3-1}\exp\left(-b_1\phi -b_2\mu - b_3\alpha \right).
\end{equation}

From the product of the likelihood function (\ref{verogg1}) and the prior distribution (\ref{pgammagg}), the joint posterior distribution for $\phi$, $\mu$ and $\alpha$ is given by,
\begin{equation}\label{postgamagg1}
\begin{aligned}
p_G\left(\phi,\mu,\alpha|\boldsymbol{t}\right)\propto &\frac{\alpha^{n+a_3-1}\mu^{n\alpha\phi+a_2-1}\phi^{a_1-1}}{\Gamma(\phi)^n}\left\{\prod_{i=1}^n{t_i^{\alpha\phi-1}}\right\}\exp\left\{ -\mu^{\alpha}\sum_{i=1}^n t_i^\alpha\right\}\times \\& \times\exp\left\{-b_1\phi -b_2\mu - b_3\alpha\right\} . 
\end{aligned}
\end{equation}

The conditional posterior distributions for $\phi,\mu$ and $\alpha$ needed for the Gibbs sampling algorithm are given as follows:
\begin{equation*}\label{postgamac1}
p_G\left(\alpha|\phi,\mu,\boldsymbol{t}\right)\propto\alpha^{n+a_3-1}\mu^{n\alpha\phi+a_2-1}\left\{\prod_{i=1}^n{t_i^{\alpha\phi-1}}\right\}\exp\left\{-b_3\alpha-\mu^{\alpha}\sum_{i=1}^n t_i^\alpha\right\},
\end{equation*}
\begin{equation}\label{postgamac2}
p_G\left(\phi|\mu,\alpha,\boldsymbol{t}\right)\propto\frac{\mu^{n\alpha\phi+a_2-1}\phi^{a_1-1}}{\Gamma(\phi)^n}\left\{\prod_{i=1}^n{t_i^{\alpha\phi-1}}\right\}\exp\left\{-b_1\phi\right\},
\end{equation}
\begin{equation*}\label{postgamac3}
p_G\left(\mu|\phi,\alpha,\boldsymbol{t}\right)\propto \mu^{n\alpha\phi+a_2-1}\exp\left\{-b_2\mu-\mu^{\alpha}\sum_{i=1}^n t_i^\alpha\right\}. 
\end{equation*}

Considering the presence of right censored observations, the joint posterior distribution for $\phi,\mu$ and $\alpha$, is proportional to the product of the likelihood function (\ref{verossccens}) and the prior distribution (\ref{pgammagg}), resulting in
\begin{equation}\label{postgamacen1}
\begin{aligned}
p_G\left(\phi,\mu,\alpha|\boldsymbol{t},\boldsymbol{\delta}\right)\propto &\frac{\alpha^{d+a_3-1}\mu^{d\alpha\phi+a_2-1}\phi^{a_1-1}}{\Gamma(\phi)^n}\left\{\prod_{i=1}^n{t_i^{{\delta_i}(\alpha\phi-1)}}\right\}\prod_{i=1}^n\left(\Gamma[\phi,(\mu t_i)^\alpha]\right)^{1-\delta_i}\times \\ & \times\exp\left\{ -b_1\phi -b_2\mu - b_3\alpha -\mu^{\alpha}\sum_{i=1}^n {\delta_i}t_i^\alpha\right\}. 
\end{aligned}
\end{equation}

The conditional posterior distributions for $\phi,\mu$ and $\alpha$ needed for the Gibbs sampling algorithm are given as follows: 
\begin{equation*}\label{postunifgamc2}
\begin{aligned}
p_G\left(\alpha|\phi,\mu,\boldsymbol{t},\boldsymbol{\delta}\right)\propto & \ \alpha^{d+a_3-1}\mu^{d\alpha\phi+a_2-1}\left\{\prod_{i=1}^n{t_i^{{\delta_i}(\alpha\phi)}}\right\}\prod_{i=1}^n\left(\Gamma[\phi,(\mu t_i)^\alpha]\right)^{1-\delta_i}\times \\ & \times\exp\left\{- b_3\alpha -\mu^{\alpha}\sum_{i=1}^n {\delta_i}t_i^\alpha\right\},
\end{aligned}
\end{equation*}
\begin{equation}\label{postunifgamc3}
\begin{aligned}
p_G\left(\phi|\mu,\alpha,\boldsymbol{t},\boldsymbol{\delta}\right)\propto& \frac{\mu^{d\alpha\phi+a_2-1}\phi^{a_1-1}}{\Gamma(\phi)^n}\left\{\prod_{i=1}^n{t_i^{{\delta_i}(\alpha\phi)}}\right\}\prod_{i=1}^n\left(\Gamma[\phi,(\mu t_i)^\alpha]\right)^{1-\delta_i}e^{-b_1\phi},
\end{aligned}
\end{equation}
\begin{equation*}\label{postunifgamc4}
p_G\left(\mu|\phi,\alpha,\boldsymbol{t},\boldsymbol{\delta}\right)\propto\mu^{d\alpha\phi+a_2-1}\prod_{i=1}^n\left(\Gamma[\phi,(\mu t_i)^\alpha]\right)^{1-\delta_i}\exp\left\{ -b_2\mu -\mu^{\alpha}\sum_{i=1}^n {\delta_i}t_i^\alpha\right\}.
\end{equation*}

Using prior opinion of experts or from the data (empirical Bayesian methods), the hyperparameters of the gamma prior distributions can be elicited using the method of moments. Let $\lambda_i$ and $\sigma_i^2$, respectively, be the mean and variance of $\theta_i$, $\theta_i\sim \f{Gamma}(a_i,b_i)$, then
\begin{equation}\label{metmomgamma}
a_i=\dfrac{\lambda_i^2}{\sigma_i^2} \ \ \mbox{ and } \ \ b_i=\dfrac{\lambda_i}{\sigma_i^2}, \ \ \ i=1,\ldots,3. 
\end{equation}

\section{Useful equations to get initial values}\label{sec:4}

In this section, an exploratory technique is discussed to obtain good initial values for numerical procedure used to obtain the MLEs for GG distribution parameters. These initial values can also be used to elicit empirical prior distributions for the parameters (use of empirical Bayesian methods). 

Although different non-informative prior distributions could be considered for the GG distribution as the Jeffreys' prior or the reference prior \cite{kass1996selection}, such priors involve transcendental function in $\phi$ such as digamma and trigamma functions with does not allow us to obtain closed-form estimator for $\phi$. A simple non-informative prior distribution representing the lack of information on the parameters can be obtained by using Jeffreys' rule \cite{kass1996selection}. As the GG distribution parameters are contained in positive intervals $(0,\infty)$, using Jeffreys' rule the the following prior is obtained
\begin{equation}\label{priorrej}
\pi_R\left(\phi,\mu,\alpha\right)\propto \frac{1}{\phi\mu\alpha}.
\end{equation}

The joint posterior distribution for $\phi,\mu$ and $\alpha$, using Jeffreys' rule is proportional to the product of the likelihood function (\ref{verogg1}) and the prior distribution (\ref{priorrej}), resulting in,

\begin{equation}\label{postrej1}
p_R(\phi,\mu,\alpha|\boldsymbol{t})=\frac{1}{d_1(\boldsymbol{t})}\frac{\alpha^{n-1}}{\phi\Gamma(\phi)^n}\mu^{n\alpha\phi-1}\left\{\prod_{i=1}^n{t_i^{\alpha\phi-1}}\right\}\exp\left\{-\mu^{\alpha}\sum_{i=1}^n t_i^\alpha\right\}. 
\end{equation}

However, the posterior density (\ref{postrej1}) is improper, i.e.,
\begin{equation*}
d_1(\boldsymbol{t})=\int\limits\limits\limits_{\mathcal{A}}\frac{\alpha^{n-1}}{\phi\Gamma(\phi)^n}\mu^{n\alpha\phi-1}\left\{\prod_{i=1}^n{t_i^{\alpha\phi-1}}\right\}\exp\left\{-\mu^{\alpha}\sum_{i=1}^n t_i^\alpha\right\}d\boldsymbol{\theta}=\infty
\end{equation*}
where $\boldsymbol{\theta}=(\phi,\mu,\alpha)$ and $\mathcal{A}=\{(0,\infty)\times(0,\infty)\times(0,\infty)\}$ is the parameter space of $\boldsymbol{\theta}$.

\begin{proof} Since $\frac{\alpha^{n-1}}{\phi\Gamma(\phi)^n}\mu^{n\alpha\phi-1}\left\{\prod_{i=1}^n{t_i^{\alpha\phi-1}}\right\}\exp\left\{-\mu^{\alpha}\sum_{i=1}^n t_i^\alpha\right\}\geq 0$, by Tonelli theorem (see Folland \cite{folland2013real}) we have
\begin{equation*}
\begin{aligned}
d_1(\boldsymbol{t})&=\int\limits\limits\limits_{\mathcal{A}}\frac{\alpha^{n-1}}{\phi\Gamma(\phi)^n}\mu^{n\alpha\phi-1}\left\{\prod_{i=1}^n{t_i^{\alpha\phi-1}}\right\}\exp\left\{-\mu^{\alpha}\sum_{i=1}^n t_i^\alpha\right\}d\boldsymbol{\theta} \\
&= \int\limits\limits\limits_0^{\infty}\int\limits\limits\limits_0^{\infty}\int\limits\limits\limits_0^{\infty}\frac{\alpha^{n-1}}{\phi\Gamma(\phi)^n}\mu^{n\alpha\phi-1}\left\{\prod_{i=1}^n{t_i^{\alpha\phi-1}}\right\}\exp\left\{-\mu^{\alpha}\sum_{i=1}^n t_i^\alpha\right\}d\mu d\phi d\alpha \\
&=\int\limits\limits\limits_0^{\infty}\int\limits\limits\limits_0^{\infty}\alpha^{n-2}\frac{\Gamma(n\phi)}{\phi\Gamma(\phi)^n}\frac{{\left(\prod_{i=1}^n t_i\right)}^{\alpha\phi-1}}{\left(\sum_{i=1}^n t_i^\alpha\right)^{n\phi}}d\phi d\alpha \geq \int\limits\limits\limits_0^{\infty}\int\limits\limits\limits_0^{1} c_1\alpha^{n-2}\phi^{n-2}\left(\frac{{\sqrt[n]{\prod_{i=1}^n t_i^\alpha}}}{\sum_{i=1}^n t_i^\alpha}\right)^{n\phi}d\phi d\alpha \\
&= \int\limits\limits\limits_0^{\infty}c_1\alpha^{n-2}\frac{\gamma\left(n-1,n \f{q}(\alpha)\right)}{\left(n \f{q}(\alpha)\right)^{n-1}}d\alpha \geq \int\limits\limits\limits_1^{\infty}g_1\frac{1}{\alpha}d\alpha = \infty,
\end{aligned}
\end{equation*}
where $\f{q}(\alpha)=\log\left(\dfrac{\sum_{i=1}^n t_i^\alpha}{{\sqrt[n]{\prod_{i=1}^n t_i^\alpha}}}\right)>0$ and $c_1$ and $g_1$ are positive constants such that the above inequalities occur. For more details and proof of the existence of these constants, see Ramos \cite{ramos2014aspectos}.\qedhere
\end{proof}

Since the Jeffreys' rule prior returned an improper posterior, a modification in this prior was considered. This modified prior is given by 
\begin{equation}\label{priorrejmod}
\pi_M\left(\phi,\mu,\alpha\right)\propto \frac{1}{\phi\mu\alpha^{\frac{1}{2}+\frac{\alpha}{1+\alpha}}} \, \cdot
\end{equation}

The present modification had two main objectives. First was to produce a modified prior for $\pi(\alpha)$ that behave similar to $\pi(\alpha)\propto \alpha^{-1}$ (see Figure 1). Second, was to construct a joint prior distribution that yields a proper posterior distribution. 
\begin{figure}[!htb]
\centering
\includegraphics[scale=0.45]{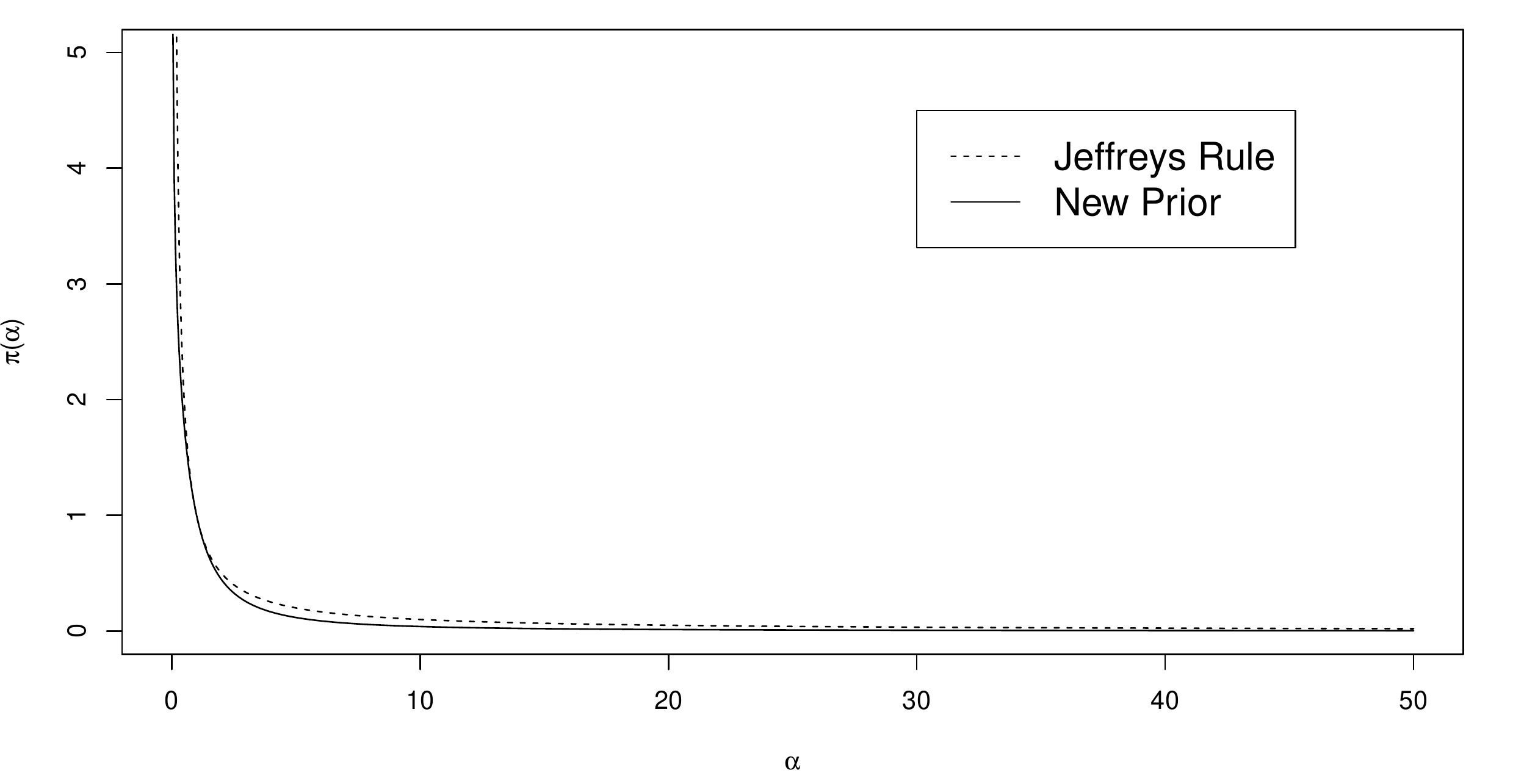}
\caption{Plot of the Jeffreys' Rule and the new Prior considering different values for $\alpha$.}\label{fsimulation2}
\end{figure}

The joint posterior distribution for $\phi,\mu$ and $\alpha$, using the prior distribution (\ref{priorrejmod}) is proportional to the product of the likelihood function (\ref{verogg1}) and the prior distribution (\ref{priorrejmod}) resulting in,
\begin{equation}\label{postrejmod1}
p_M(\phi,\mu,\alpha|\boldsymbol{t})=\frac{1}{d_2(\boldsymbol{t})}\frac{\alpha^{n-\tfrac{1}{2}-\alpha/(1+\alpha)}}{\phi\Gamma(\phi)^n}\mu^{n\alpha\phi-1}\left\{\prod_{i=1}^n{t_i^{\alpha\phi-1}}\right\}\exp\left\{-\mu^{\alpha}\sum_{i=1}^n t_i^\alpha\right\}. 
\end{equation}

This joint posterior distribution is proper, i.e.,
\begin{equation}\label{postrejmod2}
d_2(\boldsymbol{t})=\int\limits\limits\limits_{\mathcal{A}}\frac{\alpha^{n-\tfrac{1}{2}-\alpha/(1+\alpha)}}{\phi\Gamma(\phi)^n}\mu^{n\alpha\phi-1}\left\{\prod_{i=1}^n{t_i^{\alpha\phi-1}}\right\}\exp\left\{-\mu^{\alpha}\sum_{i=1}^n t_i^\alpha\right\}d\boldsymbol{\theta}<	\infty,
\end{equation}
where $\boldsymbol{\theta}=(\phi,\mu,\alpha)$ and $\mathcal{A}=\{(0,\infty)\times(0,\infty)\times(0,\infty)\}$ is the parameter space of $\boldsymbol{\theta}$. The proof of this result is presented in Appendix A at the end of the paper.

The marginal posterior distribution for $\alpha$ and $\phi$ is obtained by integrating (\ref{postrejmod1}) with respect to $\mu$, i.e.,
\begin{equation*}
p_M(\phi,\alpha|\boldsymbol{t})=\frac{1}{d_2(\boldsymbol{t})}\frac{\alpha^{n-\tfrac{1}{2}-\alpha/(1+\alpha)}}{\phi\Gamma(\phi)^n}\left\{\prod_{i=1}^n{t_i^{\alpha\phi-1}}\right\}\int\limits\limits_{0}^{\infty}\mu^{n\alpha\phi-1}\exp\left\{-\mu^{\alpha}\sum_{i=1}^n t_i^\alpha\right\}d\mu. 
\end{equation*}

From the result
\begin{equation*}
\int\limits\limits_0^{\infty} x^{\alpha-1}e^{-\beta x}dx=\frac{\Gamma(\alpha)}{\beta^\alpha}
\end{equation*}
and considering the transformation $v=\mu^\alpha$, we have
\begin{equation*}
\begin{aligned}
\int\limits\limits_0^{\infty}\mu^{n\alpha\phi-1}\exp\left\{-\mu^{\alpha}\sum_{i=1}^n t_i^\alpha\right\}d\mu &= \frac{1}{\alpha}\int\limits\limits_0^{\infty} v^{n\phi-1}\exp\left\{-v\sum_{i=1}^n t_i^\alpha\right\}dv = \dfrac{\Gamma(n\phi)}{\alpha\left(\sum_{i=1}^n t_i^\alpha\right)^{n\phi}}\ .
\end{aligned}
\end{equation*}

Therefore, the marginal posterior distribution of $\alpha$ and $\phi$ is given by
\begin{equation*}
p_M(\phi,\alpha|\boldsymbol{t})=\frac{1}{d_2(\boldsymbol{t})}\frac{\alpha^{n-\tfrac{3}{2}-\alpha/(1+\alpha)}}{\phi\Gamma(\phi)^n}\left\{\prod_{i=1}^n{t_i^{\alpha\phi-1}}\right\}\dfrac{\Gamma(n\phi)}{\left(\sum_{i=1}^n t_i^\alpha\right)^{n\phi}}\ . 
\end{equation*}

The logarithm of the above marginal distribution is given by
\begin{equation*} 
\begin{aligned}
\log\left(p_M(\alpha,\phi|\boldsymbol{t})\right)=\ &\left({n-\frac{3}{2}-\frac{\alpha}{(1+\alpha)}}\right)\log(\alpha)+\log\left(\Gamma(n\phi)\right)-\log(\phi)-\log(d_2) \\& +(\alpha\phi-1)\sum_{i=1}^n \log(t_i) -n\log\left(\Gamma(\phi)  \right) -n\phi\log\left(\sum_{i=1}^n t_i^\alpha \right).
\end{aligned}
\end{equation*}

The first-order partial derivatives of $\log\left(p_M(\alpha,\phi|\boldsymbol{t})\right)$, with respect to $\alpha$ and $\phi$, are given, respectively, by,
\begin{equation*}
\begin{aligned}
&\frac{\partial\log\left(p_M(\alpha,\phi|\boldsymbol{t})\right)}{\partial\alpha}=\kappa(\alpha)+\phi\sum_{i=1}^{n} \log(t_i) - n\phi\left(\frac{\sum_{i=1}^n t_i^\alpha \log(t_i)}{\sum_{i=1}^n t_i^\alpha} \right), \\ &\frac{\partial\log\left(p_M(\alpha,\phi|\boldsymbol{t})\right)}{\partial\phi}= n\psi(n\phi)+\alpha\sum_{i=1}^n{\log(t_i)} -\frac{1}{\phi} -n\psi(\phi)-n\log\left(\sum_{i=1}^n t_i^\alpha \right) ,
\end{aligned}
\end{equation*}
where $\kappa(\alpha)=\left({n-\frac{3}{2}-\frac{\alpha}{(1+\alpha)}}\right)\frac{1}{\alpha}-\frac{\alpha}{(1+\alpha)^2}\log(\alpha)$.
 The maximum a posteriori estimates of $\tilde{\phi}$ and $\tilde{\alpha}$ related to $p_M(\alpha,\phi|\boldsymbol{t})$ are obtained by solving
\begin{equation*}
\frac{\partial\log\left(p_M(\alpha,\phi|\boldsymbol{t})\right)}{\partial\alpha}=0 \ \ \mbox{ and }\ \ \frac{\partial\log\left(p_M(\alpha,\phi|\boldsymbol{t})\right)}{\partial\phi}=0 .
\end{equation*}

By solving $\partial\log\left(p_M(\alpha,\phi|\boldsymbol{t})\right / \partial\alpha$, we have
\begin{equation}\label{vi1}
\tilde{\phi}=\cfrac{\kappa(\tilde{\alpha})\sum_{i=1}^n t_i^{\tilde{\alpha}}}{\left(n\sum_{i=1}^{n}t_i^{\tilde{\alpha}} \log(t_i^{\tilde{\alpha}}) -\sum_{i=1}^n t_i^{\tilde{\alpha}}\sum_{i=1}^n \log(t_i^{\tilde{\alpha}}) \right)} \,\cdot
\end{equation}

By solving $\partial\log\left(p_M(\alpha,\phi|\boldsymbol{t})\right / \partial\phi$ and replacing $\phi$ by $\tilde{\phi}$ we have $\tilde{\alpha}$  can be obtained finding the minimum of
\begin{equation}\label{vi2}
h(\alpha)=n\psi\left(n\tilde{\phi}\right)+\alpha\sum_{i=1}^n \log(t_i) -\frac{1}{\tilde{\phi}}-\psi\left(\tilde{\phi}\right) -\log\left(\sum_{i=1}^n t_i^{\alpha}\right)
\end{equation}

Using one dimensional iterative methods, we obtain the value of $\tilde{\alpha}$. Replacing $\tilde{\alpha}$ in equation (\ref{vi1}), we have $\tilde{\phi}$. The preliminary estimate $\mu$ can be obtained from
\begin{equation*}
\frac{\partial\log\left(p_M(\phi,\mu,\alpha|\boldsymbol{t})\right)}{\partial\mu}=(n\tilde{\phi}\tilde{\alpha}-1)-\tilde{\alpha}\tilde{\mu}^{\tilde{\alpha}}\sum_{i=1}^n t_i^{\tilde{\alpha}}=0
\end{equation*}
and after some algebraic manipulation,
\begin{equation}\label{vi3}
\tilde{\mu}=
\begin{cases}
\left(\tfrac{n\tilde{\alpha}\tilde{\phi}-1}{\tilde{\alpha}\sum_{i=1}^{n} t_i^{\tilde{\alpha}}} \right)^{\tfrac{1}{\tilde{\alpha}}}, & \text{if } n\tilde{\alpha}\tilde{\phi}>1 \\
\left(\frac{n\tilde{\phi}}{\sum_{i=1}^{n} t_i^{\tilde{\alpha}}} \right)^{\tfrac{1}{\tilde{\alpha}}}, & \text{if }n\tilde{\alpha}\tilde{\phi}<1
\end{cases}
\end{equation}

This approach can be easily used to find good initial values $\tilde{\phi}, \tilde{\mu}$ and $\tilde{\alpha}$. These results are of great practical interest since it can be used in a Bayesian analysis to construct empirical prior distributions for $\phi, \mu$ and $\alpha$ as well as for initialization of the iterative methods to find the MLEs of the GG distribution parameters.

\section{A numerical analysis}\label{sec:5}

In this section, a simulation study via Monte Carlo method is performed in order to study the effect of the initial values obtained from (\ref{vi1}, \ref{vi2} and \ref{vi3}) in the MLEs and posterior estimates, considering complete and censored data. In addition, MCMC (Markov Chain Monte Carlo) methods are used to obtain the posterior summaries. The influence of sample sizes on the accuracy of the obtained estimates was also investigated. The following procedure was adopted in the study:
\begin{enumerate}
\item Set the values of $\boldsymbol{\theta}=(\phi,\mu,\alpha)$ and n.
\item Generate values of  the $\f{GG}(\phi,\mu,\alpha)$ distribution with size $n$.
\item Using the values obtained in step (2), calculate the estimates of $\hat{\phi},\hat{\mu}$ and $\hat{\alpha}$ using the MLEs or the Bayesian inference for the parameters $\phi, \mu$ and $\alpha$.
\item Repeat (2) and (3) $N$ times.
\end{enumerate}

The chosen values for this simulation study are $\boldsymbol{\theta}=((0.5, 0.5, 3)$,$(0.4,1.5,5))$ and $n=(50,100,200)$. The seed used to generate the random values in the R software is $2013$. The comparison between the methods was made through the calculation of the averages and standard deviations of the $N$ estimates obtained through the MLEs and posterior means. It is expected that the best estimation method has the means of the $N$ estimates closer to the true values of $\boldsymbol{\theta}$ with smaller standard deviations. It is
important to point out that, the results of this simulation study were similar for different
values of $\boldsymbol{\theta}$.

\subsection{A classical analysis}\label{sec:51}

In the lack of information on which initial values should be used in the iterative method, we generate random values such that $\tilde{\phi}\sim \f{U}(0,5)$,$\tilde{\mu}\sim \f{U}(0,5)$ and $\tilde{\alpha}\sim \f{U}(0,5)$. This procedure is denoted as ``Method 1''. On the other hand, from equations (\ref{vi1}, \ref{vi2} and \ref{vi3}), good initial values are easily obtained to be used in iterative methods to get the MLEs considering complete observations. In the presence of censored observations we discuss the possibility of using only the complete observations, in order to obtain good initial values $\tilde{\phi},\tilde{\mu}$ and $\tilde{\alpha}$. This procedure is denoted as ``Method 2''.

Tables 1-3 present the means and standard deviations of the estimates of $N=50,000$ samples obtained using the MLEs, calculated by the methods 1 and 2 for different values of $\boldsymbol{\theta}$ and $n$, using complete and censored observations. We also have the coverage probability (CP) with a nominal $95\%$ confidence level.
\begin{table}[!htb]
\centering
\caption{Estimates of the means and standard deviations of MLE's and coverage probabilities with a nominal $95\%$ confidence level obtained from $50,000$ samples using Method 1 for different values of $\boldsymbol{\theta}$ and $n$, using complete observations as well as censored observations ($20\%$ censoring).}
{\footnotesize
\begin{tabular}{ c|c|c|c|c|c|c }
  \hline
  \multicolumn{1}{c|}{$\boldsymbol{\theta}$}  & \multicolumn{3}{c|}{Complete observations} & \multicolumn{3}{c}{Censored observations} \\ 
  \hline
	& \multicolumn{1}{c|}{$n=50$} & \multicolumn{1}{c|}{$n=100$}  & \multicolumn{1}{c|}{$n=200$}  & \multicolumn{1}{c|}{$n=50$}  & \multicolumn{1}{c|}{$n=100$}  & \multicolumn{1}{c}{$n=200$} \\
	\hline
	& Mean(SD) & Mean(SD) & Mean(SD) & Mean(SD) & Mean(SD) & Mean(SD) \\
	\multicolumn{1}{c|}{ } & \multicolumn{1}{c|}{MLE}    &  \multicolumn{1}{c|}{MLE}      & \multicolumn{1}{c|}{MLE}   &  \multicolumn{1}{c|}{MLE}      & \multicolumn{1}{c|}{MLE}   &  \multicolumn{1}{c}{MLE}  \\
	\hline
  \multicolumn{1}{c|}{$\phi=0.5$}      & 1.940(1.416)	& 1.806(1.335)	& 1.788(1.248)	& 1.105(1.323) & 1.026(1.202) & 0.981(1.149) \\
  \multicolumn{1}{c|}{$\mu=0.5$}       & 1.816(1.973)	& 1.692(1.850)	& 1.646(1.821)	& 1.112(1.983) & 1.033(1.876) & 0.997(1.835) \\
  \multicolumn{1}{c|}{$\alpha=3$}      & 1.827(1.260)	& 1.821(1.061)	& 1.742(0.921)	& 3.565(2.863) & 3.254(2.218) & 3.091(1.834) \\
	\hline
  \multicolumn{1}{c|}{$\phi=0.4$}       & 1.094(0.968) & 1.050(0.957) & 0.983(0.970)	& 0.667(0.882) & 0.659(0.936) & 0.535(0.544) \\
	\multicolumn{1}{c|}{$\mu=1.5$}        & 2.318(1.724) & 2.386(3.729) & 2.352(4.507) 	& 1.842(1.617) & 1.872(2.110) & 1.608(0.894) \\
  \multicolumn{1}{c|}{$\alpha=5$}       & 3.548(2.003) & 3.539(4.282) & 3.526(1.492)	& 6.266(4.290) & 5.686(3.180) & 5.379(2.613) \\
	\hline
		& \multicolumn{1}{c|}{CP} & \multicolumn{1}{c|}{CP}  & \multicolumn{1}{c|}{CP}  & \multicolumn{1}{c|}{CP}  & \multicolumn{1}{c|}{CP}  & \multicolumn{1}{c}{CP} \\
	\hline
  \multicolumn{1}{c|}{$\phi=0.5$}        & 97.96\% & 98.17\% & 98.52\% & 70.59\% & 65.67\% & 61.67\% \\
	\multicolumn{1}{c|}{$\mu=0.5$}         & 98.31\% & 98.36\% & 98.62\% & 79.61\% & 72.29\% & 64.35\% \\
  \multicolumn{1}{c|}{$\alpha=3$}        & 44.64\% & 39.49\% & 29.58\% & 71.05\% & 67.27\% & 63.38\% \\
  \hline
	\multicolumn{1}{c|}{$\phi=0.4$}        & 98.30\% & 98.80\% & 98.80\% & 77.40\% & 74.00\% & 75.00\% \\
	\multicolumn{1}{c|}{$\mu=1.5$}         & 99.20\% & 99.00\% & 98.90\% & 83.10\% & 77.50\% & 77.20\% \\
  \multicolumn{1}{c|}{$\alpha=5$}        & 63.90\% & 52.30\% & 47.80\% & 79.40\% & 75.80\% & 76.60\% \\
  \hline
\end{tabular}
}
\end{table}


\begin{table}[!htb]
\centering
\caption{Estimates of the means and standard deviations of MLE's and coverage probabilities obtained from $50,000$ samples using Method 2 for different values of $\boldsymbol{\theta}$ and $n$, using complete observations as well as censored observations ($20\%$ censoring).}
{\footnotesize
\begin{tabular}{ r|r|r|r|r|r|r }
  \hline
  \multicolumn{1}{c|}{$\boldsymbol{\theta}$}  & \multicolumn{3}{c|}{Complete observations} & \multicolumn{3}{c}{Censored observations} \\ 
  \hline
	& \multicolumn{1}{c|}{$n=50$} & \multicolumn{1}{c|}{$n=100$}  & \multicolumn{1}{c|}{$n=200$}  & \multicolumn{1}{c|}{$n=50$}  & \multicolumn{1}{c|}{$n=100$}  & \multicolumn{1}{c}{$n=200$} \\
	\hline
	& Mean(SD) & Mean(SD) & Mean(SD) & Mean(SD) & Mean(SD) & Mean(SD) \\
	\multicolumn{1}{c|}{ } & \multicolumn{1}{c|}{IV}    &  \multicolumn{1}{c|}{IV}      & \multicolumn{1}{c|}{IV}   &  \multicolumn{1}{c|}{IV}      & \multicolumn{1}{c|}{IV}   &  \multicolumn{1}{c}{IV}  \\
	\hline
  \multicolumn{1}{c|}{$\phi=0.5$}      & 0.675(0.211)	& 0.686(0.141)	& 0.689(0.092) 	& 0.759 0.793 & 0.599 0.429	& 0.532(0.223)  \\
  \multicolumn{1}{c|}{$\mu=0.5$}       & 0.572(0.124)	& 0.571(0.069)	& 0.570(0.043)	& 0.713 1.148 & 0.565 0.399	& 0.519(0.101) \\
  \multicolumn{1}{c|}{$\alpha=3$}      & 2.576(0.432)	& 2.493(0.316)	& 2.450(0.224)	& 3.282 1.616 & 3.296 1.283	& 3.201(0.895)  \\
	\hline
  \multicolumn{1}{c|}{$\phi=0.4$}       & 0.646 0.157	& 0.661(0.105)	& 0.674 0.071	& 0.752 0.716 & 0.543(0.312)	& 0.456(0.157) \\
	\multicolumn{1}{c|}{$\mu=1.5$}        & 1.695 0.170	& 1.703(0.110)	& 1.711 0.073	& 1.876 1.225 & 1.629(0.333)	& 1.549(0.131) \\
  \multicolumn{1}{c|}{$\alpha=5$}       & 3.742 0.564	& 3.619(0.410)	& 3.538 0.293	& 4.204 1.602 & 4.603(1.299)	& 4.867(1.012) \\
	\hline
	& Mean(SD) & Mean(SD) & Mean(SD) & Mean(SD) & Mean(SD) & Mean(SD) \\
	\multicolumn{1}{c|}{ } & \multicolumn{1}{c|}{MLE}    &  \multicolumn{1}{c|}{MLE}      & \multicolumn{1}{c|}{MLE}   &  \multicolumn{1}{c|}{MLE}      & \multicolumn{1}{c|}{MLE}   &  \multicolumn{1}{c}{MLE}  \\
	\hline
  \multicolumn{1}{c|}{$\phi=0.5$}        & 0.651(0.592)	& 0.557(0.327)	& 0.523(0.185) 	& 0.738(0.778) & 0.585(0.410)	& 0.525(0.215)  \\
	\multicolumn{1}{c|}{$\mu=0.5$}         & 0.605(0.368)	& 0.537(0.158)	& 0.514(0.073) 	& 0.673(1.107) & 0.538(0.370)	& 0.499(0.086)  \\
  \multicolumn{1}{c|}{$\alpha=3$}        & 3.771(2.135)	& 3.380(1.349)	& 3.160(0.779)	& 3.523(1.842) & 3.500(1.419)	& 3.378(0.967)  \\
  \hline
  \multicolumn{1}{c|}{$\phi=0.4$}       & 0.571(0.503)  & 0.500(0.258)  & 0.491(0.152)  & 0.737(0.704) & 0.539(0.302)	& 0.457(0.154)  \\
	\multicolumn{1}{c|}{$\mu=1.5$}        & 1.681(0.517)	& 1.591(0.234)	& 1.574(0.123)	& 1.818(1.179) & 1.593(0.302)	& 1.523(0.121) \\
  \multicolumn{1}{c|}{$\alpha=5$}       & 5.563(2.763)	& 4.987(1.676)	& 4.688(1.204)	& 4.438(1.787) & 4.826(1.425)	& 5.071(1.083) \\
  \hline
\end{tabular}
}
\end{table}

\begin{table}[!htb]
\centering
\caption{$95\%$ confidence level obtained from $50,000$ samples using Method 2 for different values of $\boldsymbol{\theta}$ and $n$, using complete observations as well as censored observations ($20\%$ censoring).}
{\footnotesize
\begin{tabular}{ r|r|r|r|r|r|r }
  \hline
  \multicolumn{1}{c|}{$\boldsymbol{\theta}$}  & \multicolumn{3}{c|}{Complete observations} & \multicolumn{3}{c}{Censored observations} \\ 
  \hline
	& \multicolumn{1}{c|}{$n=50$} & \multicolumn{1}{c|}{$n=100$}  & \multicolumn{1}{c|}{$n=200$}  & \multicolumn{1}{c|}{$n=50$}  & \multicolumn{1}{c|}{$n=100$}  & \multicolumn{1}{c}{$n=200$} \\
	\hline
  \multicolumn{1}{c|}{$\phi=0.5$}      & 83.93\% & 88.34\% & 91.63\% & 94.19\% & 91.99\% & 91.55\% \\
  \multicolumn{1}{c|}{$\mu=0.5$}       & 86.58\% & 89.31\% & 91.81\% & 92.08\% & 88.56\% & 86.69\% \\
  \multicolumn{1}{c|}{$\alpha=3$}      & 92.95\% & 94.48\% & 94.89\% & 95.58\% & 96.30\% & 96.69\% \\
	\hline
	\multicolumn{1}{c|}{$\phi=0.4$}      & 91.57\% & 95.35\% & 95.35\% & 98.34\% & 99.01\% & 99.09\% \\
	\multicolumn{1}{c|}{$\mu=1.5$}       & 93.99\% & 96.08\% & 96.18\% & 98.53\% & 98.44\% & 97.58\% \\
  \multicolumn{1}{c|}{$\alpha=5$}      & 90.64\% & 91.85\% & 90.18\% & 92.79\% & 94.43\% & 96.07\% \\
  \hline
\end{tabular}
}
\end{table}

We can observe in the results shown in Tables 1 to 3, that the MLEs are strongly affected by the initial values. For example, using method 1, for $\mu = 0.5 and n = 50$, the average obtained from $N=50,000$ estimates of $\mu$ is equal to 1.816, with standard deviation equals to 1.973. Other problem using such methodology is that the coverage probability tends to decrease even considering large sample sizes.

Using our proposed methodology (method 2) it can be seen that in all cases the method gives good estimates for $\phi, \alpha$ and $\mu$. The coverage probabilities of parameters show that, as the sample size increase the intervals tend to $95\%$ as expected. Therefore, we can easily obtain good inferences for the parameters of the GG distribution for both, complete or censored data sets. 

\subsection{A Bayesian analysis}\label{sec:52}

From the Bayesian approach, we can obtain informative hyperparameter for the gamma prior distribution using the method of moments (\ref{metmomgamma}) with mean given by $\boldsymbol{\lambda}=(\tilde{\phi},\tilde{\mu},\tilde{\alpha})$, obtained using the equations (\ref{vi1}-\ref{vi3}). Additional,  we have assumed the variance given by $\boldsymbol{\sigma}_{\theta}^{2}=(1,1,1),\boldsymbol{\sigma}_{\theta}^{2}=(\tilde{\phi},\tilde{\mu},\tilde{\alpha})$ and $\boldsymbol{\sigma}_{\theta}^{2}=(10,10,10)$ to verify which of these values produce better results. The procedure was replicated considering $N=1,000$ for different values of the parameters.

The OpenBUGS was used to generate chains of the marginal posterior distributions of $\phi,\mu$ and $\alpha$ via MCMC methods. The code is given by
\begin{verbatim}
model<-function () {
for (i in 1:N) {
x[i] ~ dggamma(phi,mu,alpha)
}
phi ~ dgamma(phi1,phi2)
mu ~ dgamma(mi1,mi2)
alpha~ dgamma(alpha1,alpha2)
}
\end{verbatim}

In the case of censored data, we only have to change dggamma(phi,mu,alpha) in the code above by dggamma(phi,mu,alpha)I(delta[i],). The hyperparameters \textit{phi1,phi2,$\ldots$,alpha2} are obtained by our initial values. The main advantage of using the OpenBUGS is due its simplicity, here, the proposal distribution for generating the marginal distributions are defined by the program according to its structure. Therefore we only have to set the data to be fitted. For each simulated sample $8,500$ iterations were performed. As ``burn-in samples'', we discarded the $1000$ initial values.To decrease the autocorrelations bettwen the samples in each chain, the thin considered was $15$, obtaining at the end three chains of size 500, these results were used to obtain the posterior summaries for $\phi,\mu$ and $\alpha$.	The convergence of the Gibbs sampling algorithm was confirmed by the Geweke criterion \cite{geweke1991evaluating} under a $95\%$ confidence level.

To illustrated the importance of good initial values and informative priors, firstly we considered a simulation study using flat priors, i.e., prior distributions that have large variance, here we assume that $\theta_i\sim\f{Uniforme}(0,40)$,
$i=1,\ldots,3$ or $\theta_i\sim\f{Gamma}(0.01,0.01)$. Table 4 displays the means, standard deviations of the posterior means from $1000$ samples obtained using the flat priors calculated for different values of  $\boldsymbol{\theta}$ and $n$.

\begin{table}[ht]
\centering
\caption{Means and standard deviations from initial values and posterior means subsequently obtained from $1000$ simulated samples with different values of $\boldsymbol{\theta}$ and $n$, using the gamma prior distribution with $\theta_i\sim\f{Uniform}(0,40)$ and $\theta_i\sim\f{Gamma}(0.01,0.01)$.}
{\footnotesize
\begin{tabular}{ r|r|r|r|r|r|r }
\hline
  & \multicolumn{3}{c|}{$\theta_i\sim\f{Uniform}(0,40)$} & \multicolumn{3}{c}{$\theta_i\sim\f{Gamma}(0.01,0.01)$} \\
	\hline
  \multicolumn{1}{c|}{$\boldsymbol{\theta}$} & \multicolumn{1}{c|}{$n=50$} & \multicolumn{1}{c|}{$n=100$} & \multicolumn{1}{c|}{$n=200$} & \multicolumn{1}{c|}{$n=50$} & \multicolumn{1}{c|}{$n=100$} & \multicolumn{1}{c}{$n=200$}\\ 
	\hline
  \multicolumn{1}{c|}{$\phi=0.5$}       & 2.945(2.255) & 1.264(1.325) & 0.669(0.409) & 1.052(0.934) & 0.719(0.499) & 0.580(0.222) \\
  \multicolumn{1}{c|}{$\mu=0.5$}        & 7.454(6.442) & 2.114(3.490) & 0.646(0.708) & 1.785(2.534) & 0.772(1.079) & 0.549(0.193) \\
  \multicolumn{1}{c|}{$\alpha=3$}       & 3.465(3.673) & 3.141(1.740) & 3.033(0.867) & 8.762(44.485) & 3.955(8.192) & 3.177(0.916) \\
	\hline
	\multicolumn{1}{c|}{$\phi=0.4$}       & 1.807(1.833) & 0.773 0.813) & 0.488(0.200) & 0.864(0.845) & 0.548(0.363) & 0.452(0.168) \\
	\multicolumn{1}{c|}{$\mu=1.5$}        & 5.246(5.029) & 2.210 2.073) & 1.587(0.214) & 2.634(2.582) & 1.730(0.853) & 1.551(0.143) \\
  \multicolumn{1}{c|}{$\alpha=5$}       & 6.177(4.707) & 5.500 2.853) & 5.211(1.536) & 9.773(11.967) & 6.348(4.353) & 5.426(1.680) \\
  \hline
\end{tabular}
}
\end{table}

From this table, we observe that flat priors with vague information may affect the posterior estimates, specially for small sample sizes, for instance in the case of $n=50$ and $\alpha=3.0$ we obtained the mean of the estimates given by 8.762 with standard deviation 44.485 which is undesirable. On the other hand,  tables 5-6 display the means, standard deviations and the coverage probabilities of the estimates of $1000$ samples obtained using the  initial values (IV) and the Bayesian estimates (BE) of the posterior means, calculated for different values of  $\boldsymbol{\theta}$ and $n$, using complete data. 
\begin{table}[!htb]
\centering
\caption{Coverage probabilities with a confidence level of $95\%$ of the estimates of posterior means obtained from $1000$ simulated samples of size $n=(50,100,200)$, with different values of $\boldsymbol{\theta}$, using the gamma prior distribution with $\boldsymbol{\lambda}=(\tilde{\phi},\tilde{\mu},\tilde{\alpha}), \boldsymbol{\sigma}_{\theta}^{2}=(1,1,1),\boldsymbol{\sigma}_{\theta}^{2}=(\tilde{\phi},\tilde{\mu},\tilde{\alpha})$ and $\boldsymbol{\sigma}_{\theta}^{2}=(10,10,10)$ and complete observations.}
\resizebox{14cm}{!} {
{\footnotesize
\begin{tabular}{ r|r|r|r|r|r|r|r|r|r }
  \hline
  \multicolumn{1}{c|}{$\boldsymbol{\theta}$}  & \multicolumn{3}{c|}{$\boldsymbol{\sigma}_{\theta}^{2}=(1,1,1)$} & \multicolumn{3}{c|}{$\boldsymbol{\sigma}_{\theta}^{2}=(\tilde{\phi},\tilde{\mu},\tilde{\alpha})$} & \multicolumn{3}{c}{$\boldsymbol{\sigma}_{\theta}^{2}=(10,10,10)$} \\ 
  \hline
	& \multicolumn{1}{c|}{$n=50$} & \multicolumn{1}{c|}{$n=100$}  & \multicolumn{1}{c|}{$n=200$}  & \multicolumn{1}{c|}{$n=50$}  & \multicolumn{1}{c|}{$n=100$}  & \multicolumn{1}{c|}{$n=200$} & \multicolumn{1}{c|}{$n=50$}  & \multicolumn{1}{c|}{$n=100$}  & \multicolumn{1}{c}{$n=200$} \\
	\hline
  \multicolumn{1}{c|}{$\phi=0.5$}      & 94.00\%	& 94.60\%	& 95.50\%	& 95.30\%	& 95.80\%	& 95.40\%	& 96.60\%	& 95.50\%	& 95.30\% \\
  \multicolumn{1}{c|}{$\mu=0.5$}       & 93.00\%	& 94.30\%	& 96.20\%	& 94.90\%	& 95.40\%	& 95.20\%	& 95.20\%	& 95.60\%	& 95.60\% \\
  \multicolumn{1}{c|}{$\alpha=3$}      & 94.90\%	& 94.00\%	& 96.00\%	& 95.70\%	& 95.00\%	& 94.80\%	& 96.00\%	& 95.30\%	& 95.40\% \\
	\hline
	\multicolumn{1}{c|}{$\phi=0.4$}       & 89.30\%	& 88.20\%	& 89.90\%	& 93.70\%	& 94.40\%	& 93.70\% & 95.80\%	& 94.90\%	& 94.90\% \\
	\multicolumn{1}{c|}{$\mu=1.5$}        & 88.90\%	& 88.10\%	& 89.90\%	& 92.40\%	& 92.60\%	& 93.20\% & 94.90\%	& 94.40\%	& 94.80\% \\
  \multicolumn{1}{c|}{$\alpha=5$}       & 86.30\%	& 85.80\%	& 88.80\%	& 93.70\%	& 93.70\%	& 93.20\% & 96.60\%	& 95.30\%	& 94.60\% \\
  \hline
\end{tabular}
}
}
\end{table}

\begin{table}[!htb]
\centering
\caption{Means and standard deviations from initial values and posterior means subsequently obtained from $1000$ simulated samples with different values of $\boldsymbol{\theta}$ and $n$, using the gamma prior distribution with $\boldsymbol{\lambda}=(\tilde{\phi},\tilde{\mu},\tilde{\alpha}), \boldsymbol{\sigma}_{\theta}^{2}=(1,1,1),\boldsymbol{\sigma}_{\theta}^{2}=(\tilde{\phi},\tilde{\mu},\tilde{\alpha})$ and $\boldsymbol{\sigma}_{\theta}^{2}=(10,10,10)$ assuming complete observations}
{\footnotesize
\begin{tabular}{ r|r|r|r|r|r|r }
  \hline
  \multicolumn{1}{c|}{$\boldsymbol{\theta}$}& \multicolumn{2}{c|}{$\boldsymbol{\sigma}_{\theta}^{2}=(1,1,1)$} & \multicolumn{2}{c|}{$\boldsymbol{\sigma}_{\theta}^{2}=(\tilde{\phi},\tilde{\mu},\tilde{\alpha})$} & \multicolumn{2}{c}{$\boldsymbol{\sigma}_{\theta}^{2}=(10,10,10)$} \\ 
  \hline
	$n=50$ & Mean(SD) & Mean(SD) & Mean(SD) & Mean(SD) & Mean(SD) & Mean(SD) \\
	\multicolumn{1}{c|}{ } & \multicolumn{1}{c|}{IV}    &  \multicolumn{1}{c|}{BE}      & \multicolumn{1}{c|}{IV}   &  \multicolumn{1}{c|}{BE}      & \multicolumn{1}{c|}{IV}   &  \multicolumn{1}{c}{BE}  \\
	\hline
  \multicolumn{1}{c|}{$\phi=0.5$}      & 0.674(0.184)	& 0.741(0.300)	& 0.676(0.193)	& 0.729(0.327)	& 0.667(0.190)	& 0.893(0.603) \\
  \multicolumn{1}{c|}{$\mu=0.5$}       & 0.572(0.097)	& 0.624(0.166)	& 0.572(0.101)	& 0.623(0.177)	& 0.568(0.100)	& 0.879(0.611) \\
  \multicolumn{1}{c|}{$\alpha=3$}      & 2.572(0.439)	& 2.767(0.634)	& 2.567(0.442)	& 2.983(0.858)	& 2.592(0.432)	& 3.423(1.466) \\
	\hline
	\multicolumn{1}{c|}{$\phi=0.4$}       & 0.629(0.135) & 0.641(0.212) & 0.629(0.137) & 0.636(0.264) & 0.625(0.128) & 0.706(0.416) \\
	\multicolumn{1}{c|}{$\mu=1.5$}        & 1.683(0.142) & 1.701(0.205) & 1.684(0.145) & 1.712(0.256) & 1.678(0.138) & 1.837(0.481) \\
  \multicolumn{1}{c|}{$\alpha=5$}       & 3.772(0.551) & 4.004(0.697) & 3.767(0.554) & 4.454(1.082) & 3.779(0.545) & 4.981(1.630) \\
	\hline
	$n=100$ & Mean(SD) & Mean(SD) & Mean(SD) & Mean(SD) & Mean(SD) & Mean(SD) \\
	\multicolumn{1}{c|}{ } & \multicolumn{1}{c|}{IV}    &  \multicolumn{1}{c|}{BE}      & \multicolumn{1}{c|}{IV}   &  \multicolumn{1}{c|}{BE}      & \multicolumn{1}{c|}{IV}   &  \multicolumn{1}{c}{PD}  \\
	\hline
  \multicolumn{1}{c|}{$\phi=0.5$}       & 0.684(0.126)	& 0.678(0.240)	& 0.685(0.129)	& 0.661(0.258)	& 0.679(0.125)	& 0.690(0.372) \\
	\multicolumn{1}{c|}{$\mu=0.5$}        & 0.569(0.060)	& 0.583(0.117)	& 0.569(0.061)	& 0.578(0.124)	& 0.568(0.062)	& 0.635(0.290) \\
  \multicolumn{1}{c|}{$\alpha=3$}       & 2.489(0.304)	& 2.805(0.568)	& 2.495(0.311)	& 2.981(0.747)	& 2.502(0.307)	& 3.238(1.086) \\
  \hline
	\multicolumn{1}{c|}{$\phi=0.4$}       & 0.636(0.099) & 0.590(0.166) & 0.635(0.095) & 0.555(0.204) & 0.635(0.098) & 0.552(0.277) \\
	\multicolumn{1}{c|}{$\mu=1.5$}        & 1.680(0.102) & 1.652(0.155) & 1.680(0.098) & 1.634(0.186) & 1.679(0.100) & 1.646(0.278) \\
  \multicolumn{1}{c|}{$\alpha=5$}       & 3.698(0.422) & 4.143(0.650) & 3.701(0.424) & 4.650(1.052) & 3.707(0.417) & 5.112(1.475) \\
  \hline
	$n=200$& Mean(SD) & Mean(SD) & Mean(SD) & Mean(SD) & Mean(SD) & Mean(SD) \\
	\multicolumn{1}{c|}{ } & \multicolumn{1}{c|}{IV}    &  \multicolumn{1}{c|}{BE}      & \multicolumn{1}{c|}{IV}   &  \multicolumn{1}{c|}{BE}      & \multicolumn{1}{c|}{IV}   &  \multicolumn{1}{c}{BE}  \\
	\hline
  \multicolumn{1}{c|}{$\phi=0.5$}       & 0.687(0.084)	& 0.605(0.169)	& 0.688(0.086)	& 0.593(0.185)	& 0.685(0.084)	& 0.582(0.216) \\
	\multicolumn{1}{c|}{$\mu=0.5$}        & 0.569(0.038)	& 0.545(0.072)	& 0.570(0.039)	& 0.543(0.078)	& 0.569(0.038)	& 0.544(0.126) \\
  \multicolumn{1}{c|}{$\alpha=3$}       & 2.456(0.216)	& 2.889(0.498)	& 2.454(0.224)	& 2.999(0.629)	& 2.454(0.221)	& 3.116(0.750) \\
  \hline
	\multicolumn{1}{c|}{$\phi=0.4$}       & 0.639(0.062) & 0.527(0.121) & 0.639(0.060) & 0.488(0.141) & 0.639(0.061) & 0.467(0.156) \\
	\multicolumn{1}{c|}{$\mu=1.5$}        & 1.684(0.066) & 1.602(0.106) & 1.683(0.064) & 1.574(0.117) & 1.683(0.066) & 1.561(0.134) \\
  \multicolumn{1}{c|}{$\alpha=5$}       & 3.642(0.292) & 4.369(0.622) & 3.647(0.288) & 4.821(0.958) & 3.645(0.287) & 5.125(1.223) \\
  \hline
\end{tabular}
}
\end{table}

We observed from Tables 5-6 that using the values calculated from equations (\ref{vi1}-\ref{vi3}) in the method of moments (\ref{metmomgamma}) with mean $\boldsymbol{\lambda}=\boldsymbol{\tilde{\theta}}$ and with variance $\boldsymbol{\sigma}_{\theta}^{2}=\boldsymbol{\tilde{\theta}}$ to get the hyperparameter values of the gamma distribution, we obtained very good posterior summaries for $\phi,\mu$ and $\alpha$, under a Bayesian approach. 

The parameters estimation were also considered under censored observations ($20\%$ of censoring). Here, we used the same procedures described in Martinez et al. \cite{martinez2016brief} to generate the pseudo random samples under the same conditions described in the beginning of this section. Tables 7-8 display the means, standard deviations and the coverage probabilities of the estimates of $1000$ samples obtained using the  initial values and the Bayes estimates of the posterior means, calculated for different values of  $\boldsymbol{\theta}$ and $n$, using censored observations ($20\%$ censoring)

\begin{table}[!htb]
\centering
\caption{Coverage probabilities with a confidence level of $95\%$ of the estimates of posterior means obtained from $1000$ simulated samples of size $n=(50,100,200)$, with different values of $\boldsymbol{\theta}$, using the gamma prior distribution with $\boldsymbol{\lambda}=(\tilde{\phi},\tilde{\mu},\tilde{\alpha}), \boldsymbol{\sigma}_{\theta}^{2}=(1,1,1),\boldsymbol{\sigma}_{\theta}^{2}=(\tilde{\phi},\tilde{\mu},\tilde{\alpha})$ and $\boldsymbol{\sigma}_{\theta}^{2}=(10,10,10)$ and censored observations ($20\%$ of censoring).}
\resizebox{14cm}{!} {
{\footnotesize
\begin{tabular}{ r|r|r|r|r|r|r|r|r|r }
  \hline
  \multicolumn{1}{c|}{$\boldsymbol{\theta}$}  & \multicolumn{3}{c|}{$\boldsymbol{\sigma}_{\theta}^{2}=(1,1,1)$} & \multicolumn{3}{c|}{$\boldsymbol{\sigma}_{\theta}^{2}=(\tilde{\phi},\tilde{\mu},\tilde{\alpha})$} & \multicolumn{3}{c}{$\boldsymbol{\sigma}_{\theta}^{2}=(10,10,10)$} \\ 
  \hline
	& \multicolumn{1}{c|}{$n=50$} & \multicolumn{1}{c|}{$n=100$}  & \multicolumn{1}{c|}{$n=200$}  & \multicolumn{1}{c|}{$n=50$}  & \multicolumn{1}{c|}{$n=100$}  & \multicolumn{1}{c|}{$n=200$} & \multicolumn{1}{c|}{$n=50$}  & \multicolumn{1}{c|}{$n=100$}  & \multicolumn{1}{c}{$n=200$} \\
	\hline
  \multicolumn{1}{c|}{$\phi=0,5$}      & 95.10\% & 92.20\% & 94.70\% & 95.70\% & 93.50\% & 94.80\% & 96.30\% & 94.00\% & 93.80\% \\
  \multicolumn{1}{c|}{$\mu=0,5$}       & 96.40\% & 95.70\% & 96.00\% & 97.10\% & 95.70\% & 94.70\% & 96.70\% & 94.80\% & 94.40\% \\
  \multicolumn{1}{c|}{$\alpha=3$}      & 96.00\% & 94.50\% & 96.20\% & 97.20\% & 95.70\% & 95.90\% & 97.70\% & 95.20\% & 93.90\% \\
  \hline
  \multicolumn{1}{c|}{$\phi=0.4$}       & 87.80\% & 85.20\% & 86.20\% & 94.10\% & 93.70\% & 94.30\% & 96.70\% & 95.60\% & 95.50\% \\
  \multicolumn{1}{c|}{$\mu=1.5$}        & 91.10\% & 91.20\% & 93.30\% & 94.90\% & 96.00\% & 95.90\% & 96.60\% & 96.60\% & 95.40\% \\
  \multicolumn{1}{c|}{$\alpha=5$}       & 87.30\% & 88.60\% & 91.40\% & 94.80\% & 95.50\% & 96.00\% & 97.10\% & 97.20\% & 95.60\% \\
  \hline
\end{tabular}
}
}
\end{table}

\begin{table}[!htb]
\centering
\caption{Means and standard deviations from initial values and posterior means subsequently obtained from $1000$ simulated samples with different values of $\boldsymbol{\theta}$ and $n$, using the gamma prior distribution with $\boldsymbol{\lambda}=(\tilde{\phi},\tilde{\mu},\tilde{\alpha}), \boldsymbol{\sigma}_{\theta}^{2}=(1,1,1),\boldsymbol{\sigma}_{\theta}^{2}=(\tilde{\phi},\tilde{\mu},\tilde{\alpha})$ and $\boldsymbol{\sigma}_{\theta}^{2}=(10,10,10)$ assuming censored data ($20\%$ censoring).}
{\footnotesize
\begin{tabular}{ r|r|r|r|r|r|r }
  \hline
  \multicolumn{1}{c|}{$\boldsymbol{\theta}$}& \multicolumn{2}{c|}{$\boldsymbol{\sigma}_{\theta}^{2}=(1,1,1)$} & \multicolumn{2}{c|}{$\boldsymbol{\sigma}_{\theta}^{2}=(\tilde{\phi},\tilde{\mu},\tilde{\alpha})$} & \multicolumn{2}{c}{$\boldsymbol{\sigma}_{\theta}^{2}=(10,10,10)$} \\ 
  \hline
	$n=50$ & Mean(SD) & Mean(SD) & Mean(SD) & Mean(SD) & Mean(SD) & Mean(SD) \\
	\multicolumn{1}{c|}{ } & \multicolumn{1}{c|}{IV}    &  \multicolumn{1}{c|}{BE}      & \multicolumn{1}{c|}{IV}   &  \multicolumn{1}{c|}{BE}      & \multicolumn{1}{c|}{IV}   &  \multicolumn{1}{c}{BE}  \\
	\hline
  \multicolumn{1}{c|}{$\phi=0.5$}      & 0.656(0.186)	& 0.757(0.302)	& 0.662(0.196)	& 0.744(0.330)	& 0.653(0.187)	& 0.941(0.638) \\
  \multicolumn{1}{c|}{$\mu=0.5$}       & 0.564(0.096)	& 0.602(0.152)	& 0.567(0.101)	& 0.603(0.166)	& 0.563(0.100)	& 0.897(0.625) \\
  \multicolumn{1}{c|}{$\alpha=3$}      & 2.635(0.465)	& 2.840(0.636)	& 2.623(0.471)	& 3.098(0.869)	& 2.648(0.462)	& 3.657(1.573) \\
	\hline
  \multicolumn{1}{c|}{$\phi=0.4$}       & 0.622(0.159) & 0.669(0.229) & 0.624(0.165) & 0.665(0.284) & 0.617(0.148) & 0.759(0.445) \\
  \multicolumn{1}{c|}{$\mu=1.5$}        & 1.682(0.173) & 1.688(0.223) & 1.685(0.181) & 1.704(0.279) & 1.677(0.162) & 1.860(0.503) \\
  \multicolumn{1}{c|}{$\alpha=5$}       & 3.831(0.610) & 4.041(0.718) & 3.824(0.614) & 4.503(1.055) & 3.844(0.606) & 5.060(1.595) \\
	\hline
	$n=100$ & Mean(SD) & Mean(SD) & Mean(SD) & Mean(SD) & Mean(SD) & Mean(SD) \\
	\multicolumn{1}{c|}{ } & \multicolumn{1}{c|}{IV}    &  \multicolumn{1}{c|}{BE}      & \multicolumn{1}{c|}{IV}   &  \multicolumn{1}{c|}{BE}      & \multicolumn{1}{c|}{IV}   &  \multicolumn{1}{c}{BE}  \\
	\hline
  \multicolumn{1}{c|}{$\phi=0.5$}       & 0.694(0.180)	& 0.727(0.287)	& 0.694(0.178)	& 0.703(0.310)	& 0.688(0.161)	& 0.762(0.471) \\
	\multicolumn{1}{c|}{$\mu=0.5$}        & 0.575(0.094)	& 0.583(0.141)	& 0.576(0.094)	& 0.578(0.153)	& 0.572(0.080)	& 0.672(0.394) \\
  \multicolumn{1}{c|}{$\alpha=3$}       & 2.505(0.360)	& 2.843(0.606)	& 2.501(0.357)	& 3.070(0.830)	& 2.516(0.359)	& 3.447(1.338) \\
  \hline
  \multicolumn{1}{c|}{$\phi=0.4$}       & 0.632(0.102) & 0.625(0.176) & 0.629(0.099) & 0.581(0.211) & 0.629(0.098) & 0.591(0.310) \\
  \multicolumn{1}{c|}{$\mu=1.5$}        & 1.680(0.104) & 1.645(0.155) & 1.676(0.101) & 1.621(0.179) & 1.677(0.101) & 1.657(0.309) \\
  \multicolumn{1}{c|}{$\alpha=5$}       & 3.726(0.451) & 4.146(0.642) & 3.738(0.448) & 4.736(1.056) & 3.734(0.444) & 5.274(1.573) \\
  \hline
	$n=200$& Mean(SD) & Mean(SD) & Mean(SD) & Mean(SD) & Mean(SD) & Mean(SD) \\
	\multicolumn{1}{c|}{ } & \multicolumn{1}{c|}{IV}    &  \multicolumn{1}{c|}{BE}      & \multicolumn{1}{c|}{IV}   &  \multicolumn{1}{c|}{BE}      & \multicolumn{1}{c|}{IV}   &  \multicolumn{1}{c}{BE}  \\
	\hline
  \multicolumn{1}{c|}{$\phi=0.5$}       & 0.684(0.095)	& 0.626(0.188)	& 0.683(0.092)	& 0.602(0.196)	& 0.684(0.094)	& 0.597(0.246) \\
	\multicolumn{1}{c|}{$\mu=0.5$}        & 0.569(0.044)	& 0.536(0.076)	& 0.569(0.044)	& 0.528(0.078)	& 0.568(0.044)	& 0.534(0.119) \\
  \multicolumn{1}{c|}{$\alpha=3$}       & 2.466(0.249)	& 2.982(0.583)	& 2.471(0.250)	& 3.155(0.738)	& 2.467(0.249)	& 3.357(0.998) \\
  \hline
  \multicolumn{1}{c|}{$\phi=0.4$}       & 0.637(0.066) & 0.559(0.127) & 0.637(0.066) & 0.510(0.153) & 0.634(0.063) & 0.478(0.167) \\
  \multicolumn{1}{c|}{$\mu=1.5$}        & 1.681(0.070) & 1.593(0.106) & 1.681(0.070) & 1.563(0.123) & 1.678(0.069) & 1.542(0.135) \\
  \multicolumn{1}{c|}{$\alpha=5$}       & 3.659(0.314) & 4.370(0.605) & 3.658(0.307) & 4.913(0.976) & 3.665(0.308) & 5.351(1.297) \\
  \hline
\end{tabular}
}
\end{table}

\newpage

From Tables 7-8 that using the values calculated from equations (\ref{vi1}-\ref{vi3}) in the method of moments (\ref{metmomgamma}) with mean $\boldsymbol{\lambda}=\boldsymbol{\tilde{\theta}}$ and with variance $\boldsymbol{\sigma}_{\theta}^{2}=\boldsymbol{\tilde{\theta}}$ are usefull to get the hyperparameter values of the gamma distribution under censored data allowing us to obtain good posterior summaries for $\phi,\mu$ and $\alpha$, under a Bayesian approach.

\section{Real data applications}\label{sec:6}

In this section, the proposed methodology is applied using two data sets from the literature. The GG distribution is assumed to analyze these data sets and the obtained results are compared with other models such as the Weibull, Gamma and Lognormal distributions, using the Akaike information criterion ($AIC=-2l(\boldsymbol{\hat{\alpha}};\boldsymbol{x})+2k$), corrected Akaike information criterion ($AICc=AIC+(2\,k\,(k+1))/(n-k-1)$) and the Bayesian information criterion ($BIC=-2l(\boldsymbol{\hat{\alpha}};\boldsymbol{x})+k\log(n)$), where $k$ is the number of parameters to be fitted and $\boldsymbol{\hat{\alpha}}$ is the estimate of $\boldsymbol{\alpha }$.  In all cases, Bayesian approach is used to obtain the estimates of the parameters for the different distributions. Addionaly, we assume gamma priors with hyperparameters values equal to 0.1 for the parameters of the Weibull, Gamma and Lognormal distributions. For each case, $30,000$ Gibbs samples were simulated using MCMC methods taking every ${15}^{th}$ generated sample obtaining a final sample of size $2,000$ to be used to get the posterior summaries.

\subsection{Remission times of patients with cancer}\label{sec:61}

In this section, we consider a data set that represents the remission times (in months) of a random sample of 128 bladder cancer patients \cite{lee2003statistical}. This data set does not has censored values (see Table \ref{tabledata}).

\begin{table}[ht]
\caption{Remission times (in months) of a random sample of 128 bladder cancer patients.}
\centering 
\begin{tabular}{r r r r r r r r r r r} 
\hline 
0.08 & 0.20 & 0.40 & 0.50 & 0.51 & 0.81 & 0.90 & 1.05 & 1.19 & 1.26 & 1.35 \\ 
1.40 & 1.46 & 1.76 & 2.02 & 2.02 & 2.07 & 2.09 & 2.23 & 2.26 & 2.46 & 2.54 \\ 
2.62 & 2.64 & 2.69 & 2.69 & 2.75 & 2.83 & 2.87 & 3.02 & 3.25 & 3.31 & 3.36 \\ 
3.36 & 3.48 & 3.52 & 3.57 & 3.64 & 3.70 & 3.82 & 3.88 & 4.18 & 4.23 & 4.26 \\ 
4.33 & 4.34 & 4.40 & 4.50 & 4.51 & 4.87 & 4.98 & 5.06 & 5.09 & 5.17 & 5.32 \\ 
5.32 & 5.34 & 5.41 & 5.41 & 5.49 & 5.62 & 5.71 & 5.85 & 6.25 & 6.54 & 6.76 \\ 
6.93 & 6.94 & 6.97 & 7.09 & 7.26 & 7.28 & 7.32 & 7.39 & 7.59 & 7.62 & 7.63 \\ 
7.66 & 7.87 & 7.93 & 8.26 & 8.37 & 8.53 & 8.65 & 8.66 & 9.02 & 9.22 & 9.47 \\ 
9.74 & 10.06 & 10.34 & 10.66 & 10.75 & 11.25 & 11.64 & 11.79 & 11.98 & 12.02 & 12.03 \\
12.07 & 12.63 & 13.11 & 13.29 & 13.80 & 14.24 & 14.76 & 14.77 & 14.83 & 15.96 & 16.62 \\
17.12 & 17.14 & 17.36 & 18.10 & 19.13 & 20.28 & 21.73 & 22.60 & 23.63 & 25.74 & 25.82 \\
26.31 & 32.15 & 34.26 & 36.66 & 43.01 & 46.12 & 79.05\\ [0ex]	
\hline 
\end{tabular}\label{tabledata}
\end{table}

From equations (\ref{vi1}-\ref{vi3}), we have $\tilde{\phi}=0.5665,\tilde{\mu}=0.2629$ and $\tilde{\alpha}=1.4669$. Using (\ref{metmomgamma}) and assuming as mean  $\boldsymbol{\lambda}=\boldsymbol{\tilde{\theta}}$ and variance $\boldsymbol{\sigma}_{\theta}^{2}=\boldsymbol{\tilde{\theta}}$, we obtain the priors $\phi\sim\f{Gamma}(0.5665,1) ,\mu\sim\f{Gamma}(0.2629,1)$ and $\alpha\sim\f{Gamma}(1.4669,1)$. The posterior summaries obtained from MCMC methods are given in Table \cite{tablestiapli}.
\begin{table}[!ht]
\caption{Posterior medians, standard deviations and $95\%$ credibility intervals for $\phi,\mu$ and $\alpha$}
\centering 
\begin{center}
  \begin{tabular}{ c | c | c | c}
    \hline
		$\boldsymbol{\theta}$  & Median & SD & $CI_{95\%}(\theta)$ \\ \hline
    \ \ $\phi$ \ \   & 2.5775 & 0.7485  & (1.3444; 4.2175) \\ \hline
    \ \ $\mu$   \ \  & 0.5712 & 0.7215  & (0.1575; 2.5821) \\ \hline
    \ \ $\alpha$ \ \ & 0.6300 & 0.1110  & (0.4789; 0.9183) \\ \hline
  \end{tabular}\label{tablestiapli}
\end{center}
\end{table}

\begin{figure}[!htb]
\centering
\includegraphics[scale=0.5]{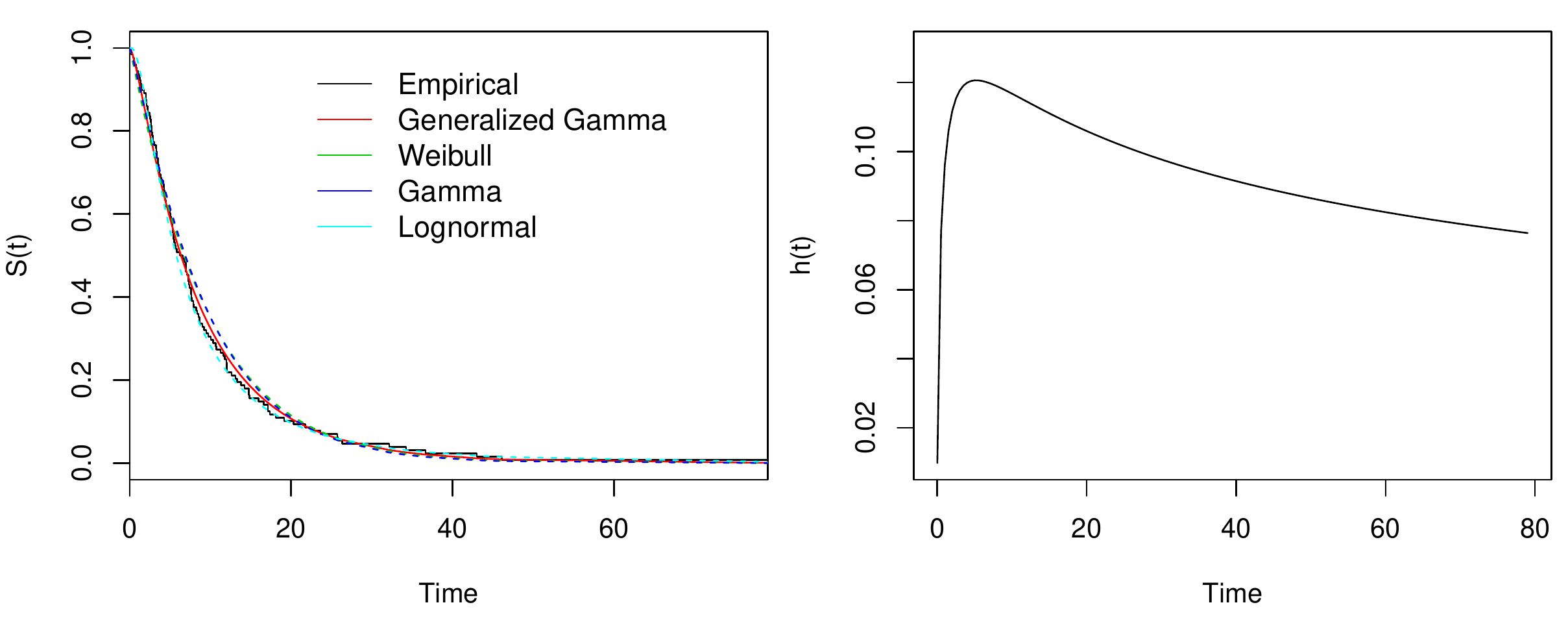}
\caption{Survival function fitted by the empirical and by different p.d.f considering the bladder cancer data set and the hazard function fitted by a GG distribution.}\label{fig2}
\end{figure}

Figure \ref{fig2} shows the survival function fitted by the different probability distributions and the empirical survival function. Table 10 presents the results of the AIC, AICc and BIC criteria for the different probability distributions, considering the blader cancer data introduced in Table 8. 

\begin{table}[!ht]
\caption{Results of AIC, AICc and BIC criteria for different probability distributions considering the bladder cancer  data set introduced in Table 8.}
\centering 
\begin{center}
  \begin{tabular}{ c | c | c | c| c}
    \hline
		Criteria & G. Gamma & \ Weibull \ & \ \ Gamma \ \ & Lognormal \\ \hline
      AIC  & \textbf{828.05} & 832.15 & 830.71 & 834.17 \\ \hline
			AICc & \textbf{828.24} & 832.25 & 830.81 & 834.27 \\ \hline
			BIC  &  836.61  & 837.86 & \textbf{836.42} & 839.88 \\ \hline
  \end{tabular}\label{tableaci1}
\end{center}
\end{table}

Based on the AIC and AICc criteria, we concluded from the results of Table \ref{tableaci1} that the GG distribution has the best fit for the bladder cancer data. 

\subsection{Data from an industrial experiment}\label{sec:62}

In this section, we considered a lifetime data set related to the cycles to failure for a batch of $60$ electrical appliances in a life test introduced by Lawless \cite{lawless2011statistical} (+ indicates the presence of censorship).
\begin{table}[!htb]
\caption{Data set of cycles to failure for a group of $60$ electrical appliances in a life test (+ indicates the presence of censorship).}
\centering 
\begin{tabular}{c c c c c c c c c c c } 
\hline 
0.014 & 0.034 & 0.059 & 0.061 & 0.069 & 0.080 & 0.123 & 0.142 & 0.165 & 0.210 \\
0.381 & 0.464 & 0.479 & 0.556 & 0.574 & 0.839 & 0.917 & 0.969 & 0.991 & 1.064 \\
1.088 & 1.091 & 1.174 & 1.270 & 1.275 & 1.355 & 1.397 & 1.477 & 1.578 & 1.649 \\
1.702 & 1.893 & 1.932 & 2.001 & 2.161 & 2.292 & 2.326 & 2.337 & 2.628 & 2.785 \\
2.811 & 2.886 & 2.993 & 3.122 & 3.248 & 3.715 & 3.790 & 3.857 & 3.912 & 4.100 \\
4.106 & 4.116 & 4.315 & 4.510 & 4.580 & 4.580+ & 4.580+ & 4.580+ & 4.580+ & 4.580+\\ [0ex] 
\hline 
\end{tabular}\label{tabledata2}
\end{table}

From equations (\ref{vi1}-\ref{vi3}), we have $\tilde{\phi}=0.5090,\tilde{\mu}=0.2746$ and $\tilde{\alpha}=1.7725$. Using (\ref{metmomgamma}) and assuming as mean  $\boldsymbol{\lambda}=\boldsymbol{\tilde{\theta}}$ and variance $\boldsymbol{\sigma}_{\theta}^{2}=\boldsymbol{\tilde{\theta}}$, we elicit the priors $\phi\sim\f{Gamma}(0.5090,1) ,\mu\sim\f{Gamma}(0.2746,1)$ and $\alpha\sim\f{Gamma}(1.7725,1)$. The posterior summaries obtained from MCMC methods are given in Table \ref{tableest2}.
\begin{table}[!ht]
\caption{Posterior medians, standard deviations and $95\%$ credibility intervals for $\phi,\mu$ and $\alpha$}
\centering 
\begin{center}
  \begin{tabular}{ c | c | c | c}
    \hline
		$\boldsymbol{\theta}$  & Median & SD & $CI_{95\%}(\theta)$ \\ \hline
    \ \ $\phi$ \ \   & 0.2689 & 0.1856  & (0.1097; 0.8193) \\ \hline
    \ \ $\mu$   \ \  & 0.1988 & 0.0632  & (0.1538; 0.3685) \\ \hline
    \ \ $\alpha$ \ \ & 2.5850 & 1.3094  & (1.0629; 5.6785) \\ \hline
  \end{tabular}\label{tableest2}
\end{center}
\end{table}

\begin{figure}[!htb]
\centering
\includegraphics[scale=0.5]{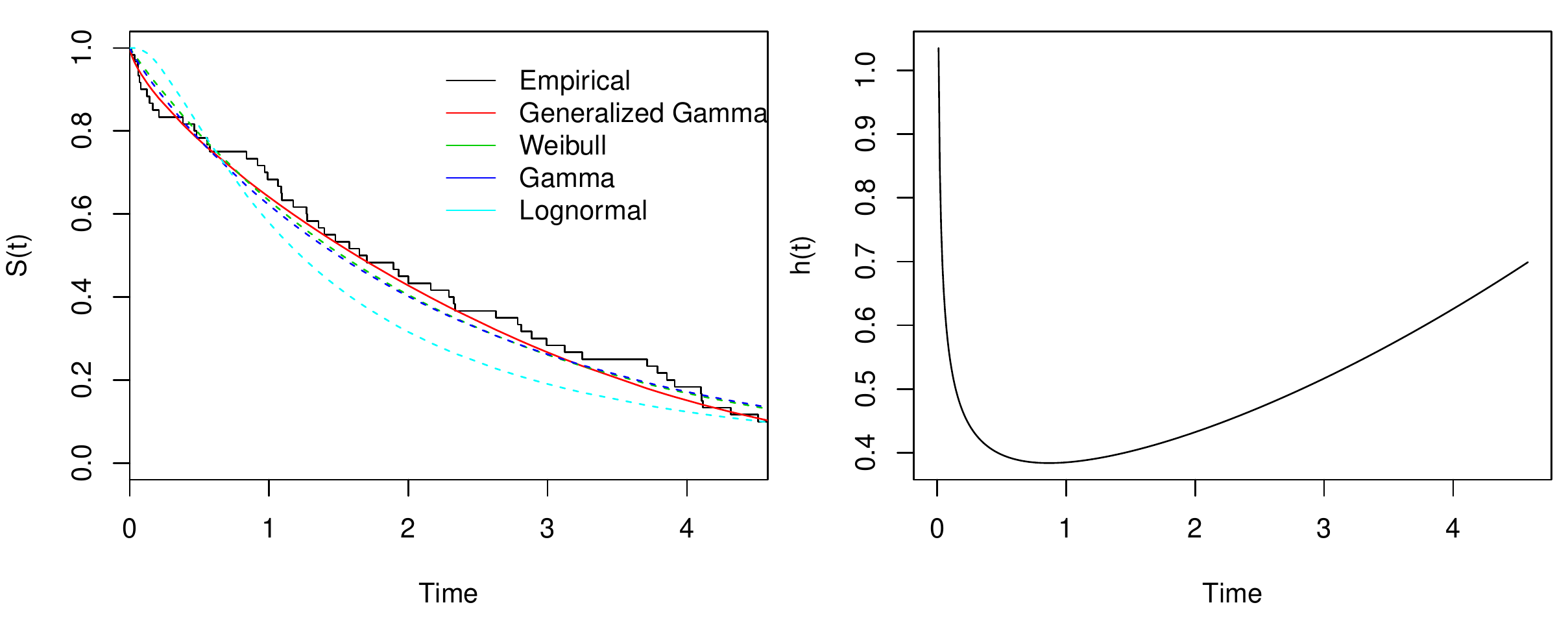}
\caption{Survival function fitted by the empirical and different p.d.f considering the data set related to the industrial data set and the hazard function fitted by a GG distribution.}\label{fig3}
\end{figure}

Figure \ref{fig3} shows the survival function fitted by probability distributions and the empirical survival function. Table \ref{aictable} presents the results of the AIC, AICc and BIC criteria for different probability distributions, considering the industrial failure data set introduced in Table \ref{tabledata2}.

\begin{table}[!ht]
\caption{Results of DIC criteria, AIC and BIC for different probability distributions considering the data of cycles to failure introduced in Table 11}
\centering 
\begin{center}
  \begin{tabular}{ c | c | c | c| c}
    \hline
		Criteria & G. Gamma & \ Weibull \ & \ \ Gamma \ \ & Lognormal \\ \hline
      AIC  & \textbf{199.19} & 202.11 & 201.80 & 216.55 \\ \hline
			AICc & \textbf{199.62} & 202.32 & 202.01 & 216.76 \\ \hline
			BIC  & \textbf{205.47} & 206.30 & 205.99 & 220.74 \\ \hline
  \end{tabular}\label{aictable}
\end{center}
\end{table}

Based on the AIC and AICc criteria, we concluded from Table \ref{aictable} that the GG distribution has the best fit for the industrial data.  Additionally, the GG distribution is the only one between probability distributions assumed that allows bathtub shape hazard. This is an important point for the use of the GG distribution in applications.

\section{Concluding remarks}\label{sec:10}

The generalized gamma distribution has played an important role as lifetime data, providing great flexibility of fit. In this paper, we showed that under the classical and Bayesian approaches, the parameter estimates usually do not exhibit stable performance in which can lead, in many cases, to different results.

This problem is overcome by proposing some empirical exploration methods aiming to obtain good initial values to be used in iterative procedures to find MLEs for the parameters of the GG distribution. These values were also used to elicit empirical prior distributions (use of empirical Bayesian methods).	These results are of great practical interest since the GG distribution can be easily used as appropriated model in different applications.

\bibliographystyle{tfs}

\bibliography{reference}

\section*{Appendix A}
\begin{proof} Since $\dfrac{\alpha^{n-\tfrac{1}{2}-\alpha/(1+\alpha)}}{\phi\Gamma(\phi)^n}\mu^{n\alpha\phi-1}\prod_{i=1}^n{t_i^{\alpha\phi-1}}\exp\left(-\mu^{\alpha}\sum_{i=1}^n t_i^\alpha\right)\geq0$ by Tonelli theorem \cite{folland2013real} we have
\begin{equation*}
\begin{aligned}
d_2&= \int\limits\limits\limits_0^\infty \int\limits\limits\limits_0^\infty \int\limits\limits\limits_0^\infty \frac{\alpha^{n-\tfrac{1}{2}-\alpha/(1+\alpha)}}{\phi\Gamma(\phi)^n}\mu^{n\alpha\phi-1}\left\{\prod_{i=1}^n{t_i^{\alpha\phi-1}}\right\}\exp\left\{-\mu^{\alpha}\sum_{i=1}^n t_i^\alpha\right\}d\mu d\phi d\alpha= \\
&=\int\limits\limits_0^{\infty}\int\limits\limits_0^{\infty}\frac{\alpha^{n-\tfrac{3}{2}-\tfrac{\alpha}{(1+\alpha)}}}{\phi\Gamma(\phi)^n}\left\{\prod_{i=1}^n{t_i^{\alpha\phi-1}}\right\}\dfrac{\Gamma(n\phi)}{\left(\sum_{i=1}^n t_i^\alpha\right)^{n\phi}}d\phi d\alpha=s_1+s_2+s_3+s_4,
\end{aligned}
\end{equation*}
where
\begin{equation*}
\begin{aligned}
s_1&=\int\limits\limits_0^{1}\int\limits\limits_0^{1}\frac{\alpha^{n-\tfrac{3}{2}-\tfrac{\alpha}{(1+\alpha)}}}{\phi\Gamma(\phi)^n}\left\{\prod_{i=1}^n{t_i^{\alpha\phi-1}}\right\}\dfrac{\Gamma(n\phi)}{\left(\sum_{i=1}^n t_i^\alpha\right)^{n\phi}}d\phi d\alpha \\
&< \int\limits\limits_0^{1}c'_1\alpha^{n-\tfrac{3}{2}-\tfrac{\alpha}{(1+\alpha)}}\frac{\gamma(n-1,n\f{q}(\alpha))}{(n\f{q}(\alpha))^{n-1}}d\alpha<\int\limits\limits_0^{1} g'_1\alpha^{n-\tfrac{3}{2}} d\alpha < \infty.
\end{aligned}
\end{equation*}
\begin{equation*}
\begin{aligned}
s_2&=\int\limits\limits_1^{\infty}\int\limits\limits_0^{1}\frac{\alpha^{n-\tfrac{3}{2}-\tfrac{\alpha}{(1+\alpha)}}}{\phi\Gamma(\phi)^n}\left\{\prod_{i=1}^n{t_i^{\alpha\phi-1}}\right\}\dfrac{\Gamma(n\phi)}{\left(\sum_{i=1}^n t_i^\alpha\right)^{n\phi}}d\phi d\alpha\\&<\int\limits\limits_1^{\infty}c'_1\alpha^{n-\tfrac{3}{2}-\tfrac{\alpha}{(1+\alpha)}}\frac{\gamma(n-1,n\f{q}(\alpha))}{(n\f{q}(\alpha))^{n-1}}d\alpha<\int\limits\limits_1^{\infty} g'_2\alpha^{-\tfrac{3}{2}} d\alpha < \infty.
\end{aligned}
\end{equation*}
\begin{equation*}
\begin{aligned}
s_3&=\int\limits\limits_0^{1}\int\limits\limits_1^{\infty}\frac{\alpha^{n-\tfrac{3}{2}-\tfrac{\alpha}{(1+\alpha)}}}{\phi\Gamma(\phi)^n}\left\{\prod_{i=1}^n{t_i^{\alpha\phi-1}}\right\}\dfrac{\Gamma(n\phi)}{\left(\sum_{i=1}^n t_i^\alpha\right)^{n\phi}}d\phi d\alpha, \\ &<\int\limits\limits_0^1 c_2a^{n-\frac{3}{2}-\frac{\alpha}{1+\alpha}}\frac{\Gamma(\frac{n-1}{2},n\f{p}(\alpha))}{(n\f{p}(\alpha))^{\frac{n-1}{2}}}d\alpha<\int\limits\limits_0^{1} g'_3\alpha^{-\tfrac{1}{2}} d\alpha<\infty.
\end{aligned}
\end{equation*}
\begin{equation*}
\begin{aligned}
s_4&=\int\limits\limits_1^{\infty}\int\limits\limits_1^{\infty}\frac{\alpha^{n-\tfrac{3}{2}-\tfrac{\alpha}{(1+\alpha)}}}{\phi\Gamma(\phi)^n}\left\{\prod_{i=1}^n{t_i^{\alpha\phi-1}}\right\}\dfrac{\Gamma(n\phi)}{\left(\sum_{i=1}^n t_i^\alpha\right)^{n\phi}}d\phi d\alpha\\&<\int\limits\limits_1^{\infty}c_2a^{n-\frac{3}{2}-\frac{\alpha}{1+\alpha}}\frac{\Gamma(\frac{n-1}{2},n\f{p}(\alpha))}{(n\f{p}(\alpha))^{\frac{n-1}{2}}}d\alpha<\int\limits\limits_1^{\infty} g'_4\alpha^{-\tfrac{3}{2}} d\alpha<\infty.
\end{aligned}
\end{equation*}
where $c'_1$, $c'_2$, $g'_1$, $g'_2$, $g'_3$ and $g'_4$ are positive constants such that the above inequalities occur. For more details and proof of the existence of these constants, see Ramos \cite{ramos2014aspectos}. Therefore, we have: $d_4=s_1+s_2+s_3+s_4<\infty$.\qedhere
\end{proof}

\end{document}